\documentclass[sigconf,natbib=false]{acmart}

\AtBeginDocument{%
  }

\setcopyright{acmlicensed}
\copyrightyear{ }
\acmYear{ }
\acmDOI{XXXXXXX.XXXXXXX}

\acmConference[Conference acronym 'XX]{Make sure to enter the correct
  conference title from your rights confirmation emai}{
   }{}
\acmISBN{ }

\RequirePackage[
  datamodel=acmdatamodel,
  style=acmnumeric,
  ]{biblatex}

\begin{document}

%%
%% The "title" command has an optional parameter,
%% allowing the author to define a "short title" to be used in page headers.
\title{  An AI-Based System Utilizing IoT-Enabled Ambient Sensors and LLMs for Complex Activity Tracking  }

%%
%% The "author" command and its associated commands are used to define
%% the authors and their affiliations.
%% Of note is the shared affiliation of the first two authors, and the
%% "authornote" and "authornotemark" commands
%% used to denote shared contribution to the research.

\author{ Yuan Sun }
\affiliation{%
  \institution{WINLAB, Rutgers University}
  \city{ Piscataway, NJ, }
  \country{ USA}}
  \email{ ys820@soe.rutgers.edu }

  \author{ Jorge Ortiz }
\affiliation{%
  \institution{WINLAB, Rutgers University}
  \city{ Piscataway, NJ, }
  \country{ USA}}
  \email{ jorge.ortiz@rutgers.edu }

%%
%% By default, the full list of authors will be used in the page
%% headers. Often, this list is too long, and will overlap
%% other information printed in the page headers. This command allows
%% the author to define a more concise list
%% of authors' names for this purpose.
\renewcommand{\shortauthors}{  }

%%
%% The abstract is a short summary of the work to be presented in the
%% article.
\begin{abstract}
Complex activity recognition plays an important role in elderly care assistance. However, the reasoning ability of edge devices is constrained by the classic machine learning model capacity. In this paper, we present a non-invasive ambient sensing system that can detect multiple activities and apply large language models (LLMs) to reason the activity sequences. This method effectively combines edge devices and LLMs to help elderly people in their daily activities, such as reminding them to take pills or handling emergencies like falls. The LLM-based edge device can also serve as an interface to interact with elderly people, especially with memory issue, assisting them in their daily lives. By deploying such a system, we believe that the smart sensing system can improve the quality of life for older people and provide more efficient protection.
\end{abstract}

%%
%% The code below is generated by the tool at http://dl.acm.org/ccs.cfm.
%% Please copy and paste the code instead of the example below.
%%
% \begin{CCSXML}
% <ccs2012>
%  <concept>
%   <concept_id>00000000.0000000.0000000</concept_id>
%   <concept_desc>Do Not Use This Code, Generate the Correct Terms for Your Paper</concept_desc>
%   <concept_significance>500</concept_significance>
%  </concept>
%  <concept>
%   <concept_id>00000000.00000000.00000000</concept_id>
%   <concept_desc>Do Not Use This Code, Generate the Correct Terms for Your Paper</concept_desc>
%   <concept_significance>300</concept_significance>
%  </concept>
%  <concept>
%   <concept_id>00000000.00000000.00000000</concept_id>
%   <concept_desc>Do Not Use This Code, Generate the Correct Terms for Your Paper</concept_desc>
%   <concept_significance>100</concept_significance>
%  </concept>
%  <concept>
%   <concept_id>00000000.00000000.00000000</concept_id>
%   <concept_desc>Do Not Use This Code, Generate the Correct Terms for Your Paper</concept_desc>
%   <concept_significance>100</concept_significance>
%  </concept>
% </ccs2012>
% \end{CCSXML}

% \ccsdesc[500]{Do Not Use This Code~Generate the Correct Terms for Your Paper}
% \ccsdesc[300]{Do Not Use This Code~Generate the Correct Terms for Your Paper}
% \ccsdesc{Do Not Use This Code~Generate the Correct Terms for Your Paper}
% \ccsdesc[100]{Do Not Use This Code~Generate the Correct Terms for Your Paper}

%%
%% Keywords. The author(s) should pick words that accurately describe
%% the work being presented. Separate the keywords with commas.

% \received{20 February 2007}
% \received[revised]{12 March 2009}
% \received[accepted]{5 June 2009}

%%
%% This command processes the author and affiliation and title
%% information and builds the first part of the formatted document.
\maketitle

\section{Introduction}
Non-intrusive sensors are crucial for modern sensing applications, particularly in fields requiring continuous and unobtrusive monitoring, such as elderly care\cite{lentzas2020non}. These sensors, which do not rely on cameras, offer significant advantages, including enhanced comfort and privacy for users. By seamlessly integrating into the environment without requiring direct interaction or visible placement, they minimize aesthetic and social intrusiveness\cite{chidurala2022iot}. Additionally, non-intrusive sensors reduce deployment and maintenance efforts, leveraging existing infrastructure and eliminating the need for frequent battery replacements\cite{yu2013performance}. This ease of installation and low maintenance makes them practical for long-term use. Furthermore, they provide flexibility and scalability in sensing applications, capturing a broad range of environmental data indirectly, thus enabling comprehensive monitoring across diverse contexts.

In this work, we build a non-intrusive smart sensing system that relies on the reasoning ability of large language models (LLMs) to assist in elderly care. The system can detect complex activities composed of more than two atomic activities. An atomic activity refers to unit-level activities that can be captured by sensors within a short time window and cannot be broken down further. Besides detecting normal atomic activities, we also use LLMs for high-level explanations and reasoning. To adapt to the memory constraints of our edge devices, we employ a local inference model on the IoT devices before sending the data to the LLM. This approach reduces the transmission burden via the wireless network, ensuring efficient and real-time processing.

\section{LLM and Complex Reasoning}
Our system first collects non-intrusive sensor data from ambient sensors. These sensors detect atomic-level activities. Due to the limited memory capacity of the sensors, a small model runs locally to perform initial inferences. Deploying a small model locally allows for real-time processing, reducing latency and ensuring immediate response to critical activities. Additionally, local processing minimizes the amount of data transmitted to the cloud, enhancing privacy and reducing bandwidth usage. Once an atomic activity is detected, it is sent to a cloud server where an LLM is deployed. The LLM applies its reasoning and interaction capabilities with the user. For instance, if the sequence of activities indicates eating and drinking without sanitizing, the system can remind the user to wash their hands. Another example is if the user drinks water but forgets to take their pills; the LLM will interact with the user to remind them to take their medication. Additionally, if a user is detected as getting dressed but skipping a critical item like a coat on a cold day, the LLM can prompt them to wear it, ensuring their well-being. Furthermore, in the event of a medical emergency, the system can recognize distress signals or calls for help, and promptly alert emergency services, providing vital assistance when needed.

\begin{figure}[h]
  \centering
  \includegraphics[width=\linewidth]{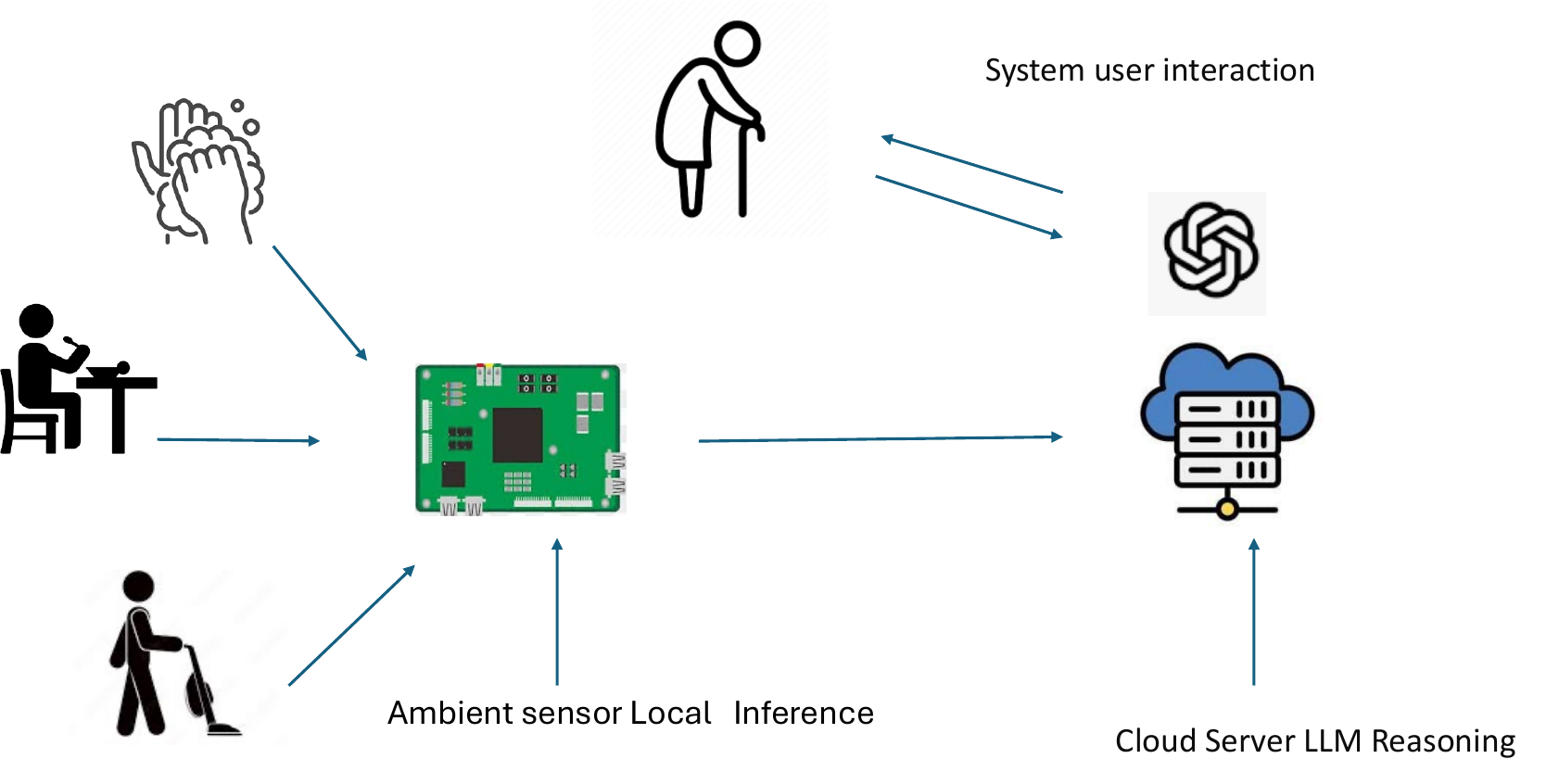}
\caption{The system employs both local inference using ambient sensors and reasoning via a cloud-based LLM. The sensors detect atomic activities, and the cloud server receives these activity sequences as context to further detect higher-level meanings, make decisions, and interact with the user.}
  \label{system}
\end{figure}

\section {sensor board setup}

Our goal is to construct a device that enhances the generalizability of sensor data explanation while safeguarding user privacy. The device (figure ~\ref{sensor_board}) , installed in the environment, utilizes non-invasive sensors to detect human activities. These sensors include PIR (motion), IMU (accelerometers), audio, RGB, pressure, humidity, magnetometer, gas, and temperature sensors. They are widely available and well-suited for generalizing smart space sensing capabilities.

Smart space sensors produce two types of outputs: binary and continuous. Binary sensors, like PIR motion sensors, provide "on" and "off" states. Although useful for indicating environmental status, they may be inadequate for analyzing complex activities~ \cite{gochoo2017dcnn, arrotta2022dexar}. Our device uses PIR sensors for motion detection, transmitting "ON" and "OFF" signals. In contrast, accelerometers offer continuous outputs, tracking movement along the X, Y, and Z axes. We utilize them to detect environmental vibrations, such as different vibration frequencies for pouring water versus running~\cite{ahmed2017improving, yan2017wearable, rawashdeh2016wearable}.

The RGB sensor detects colors within a specific range, identifying the presence of colors based on the activated channels. Pressure sensors measure barometric pressure, providing insights into weather patterns and indoor air quality, especially in HVAC operations. Magnetometers detect magnetic fields, but readings can be influenced by nearby electronic devices and human activity due to RF electromagnetic fields~\cite{martinez2004comparison, bernardi2000specific}. Gas sensors detect substances like ethanol, alcohol, and carbon monoxide, useful for monitoring air quality changes due to activities such as heavy running.

\begin{figure}[h]
  \centering
  \includegraphics[width=0.5\linewidth]{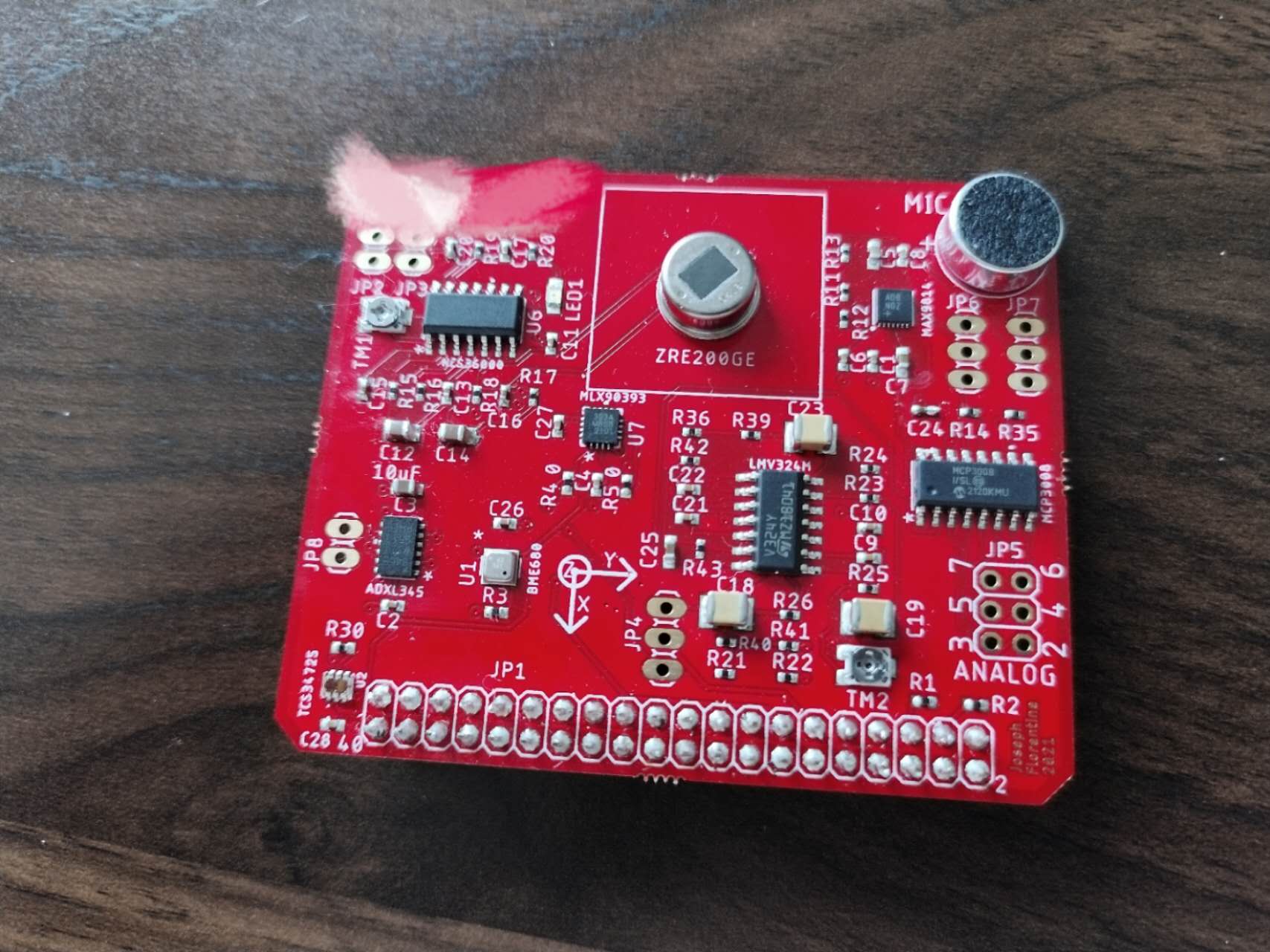}
  \caption{The non-intrusive sensor board we design for our system}
  \label{sensor_board}
\end{figure}

The Raspberry Pi Model B+ (Figure~\ref{sensor_board}) serves as the central processing unit, interfacing with a custom sensor board via GPIO pins. The sensor board, equipped with various ambient sensors, collects data such as temperature, humidity, and motion. The Raspberry Pi, powered by a Broadcom BCM2837B0 quad-core ARM Cortex-A53 64-bit SoC at 1.4GHz with 1GB LPDDR2 SDRAM, runs a Python script to read sensor values at regular intervals, process the data, and store it for further analysis. This setup, featuring dual-band 802.11ac wireless LAN, Bluetooth 4.2, Gigabit Ethernet, and multiple USB ports, enables real-time monitoring and data collection essential for elderly care assistance and activity tracking. Although our current sensor setup is insufficient to run a model for data embedding locally, we plan to implement local atomic activity detection in the next stage of this work.

\begin{figure}[h]
  \centering
  \includegraphics[width=0.5\linewidth]{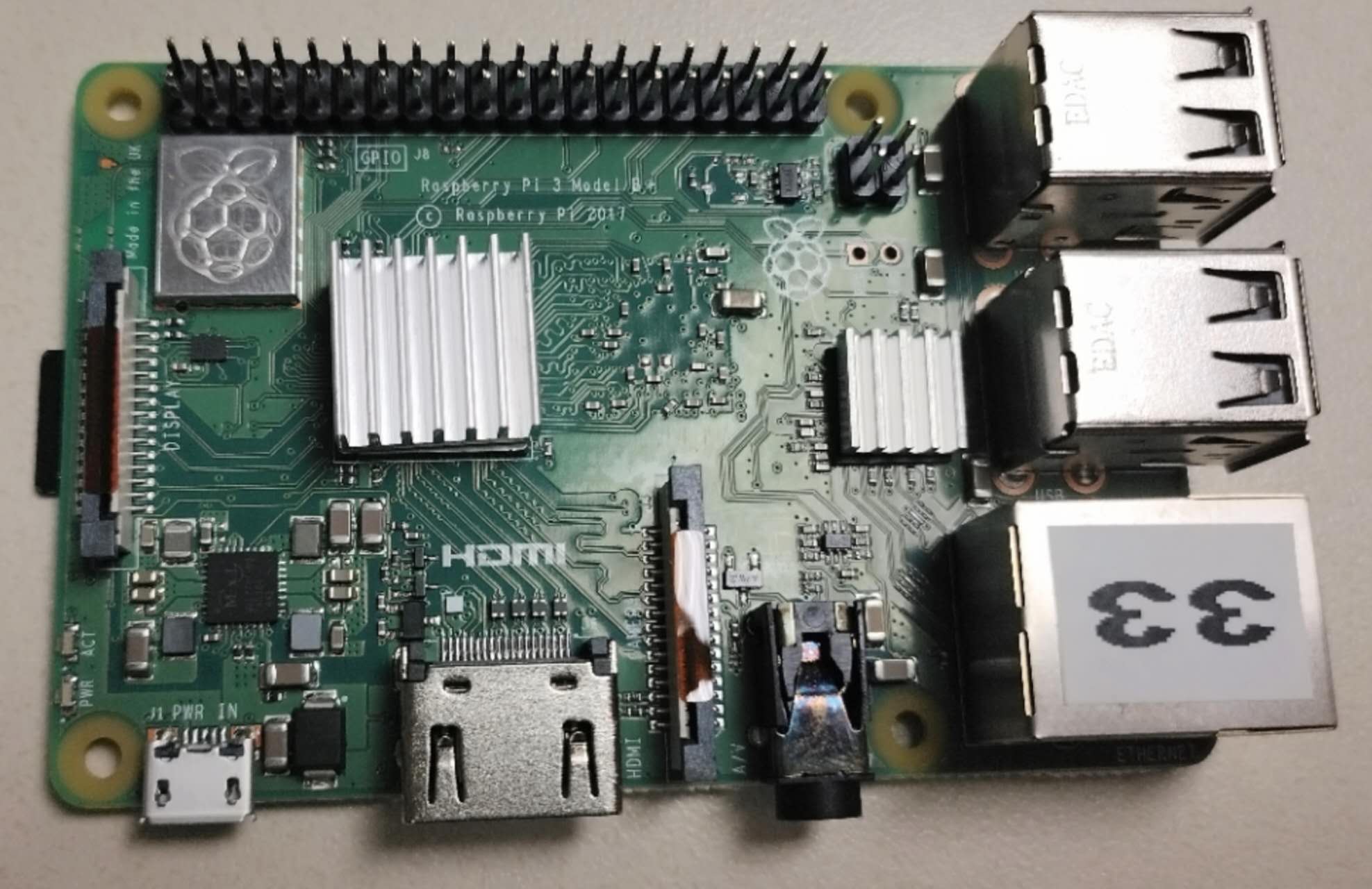}
  \caption{Raspberry Pi Model B+ used in our ambient sensor setup, facilitating seamless integration for elderly care assistance and activity tracking}
  \label{sensor_board}
\end{figure}

\section { initial experiment}

In our initial experiment, we will use the ambient sensor to collect atomic activities and test the LLM model on the cloud. The target is to determine if it possesses the reasoning capabilities that can assist in the daily lives of elderly individuals.

\subsection{Data Collection}

\begin{figure}[h]
  \centering
  \includegraphics[width=\linewidth]{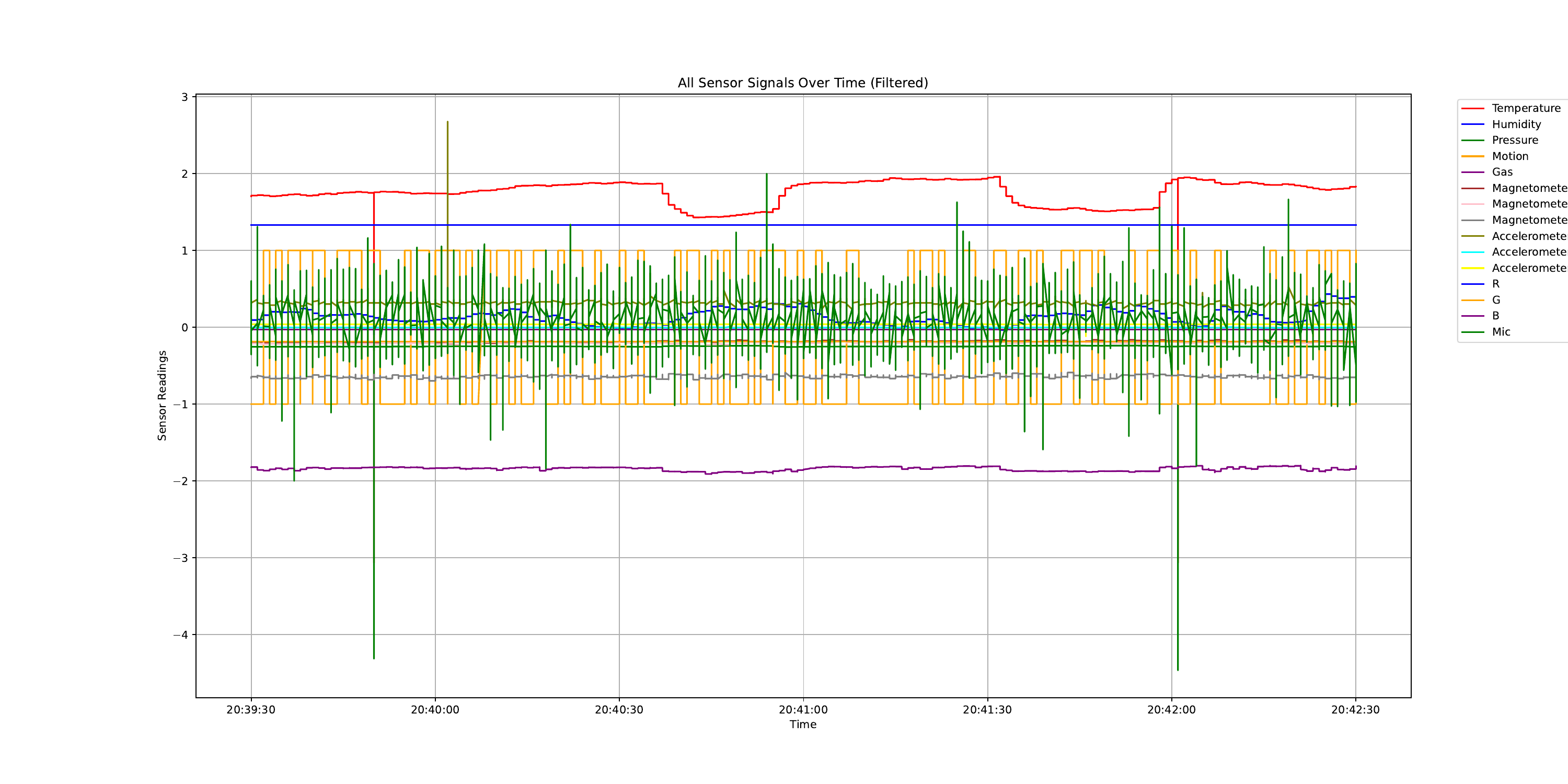}
 \caption{Eating activity collected by the ambient sensor}
  \label{data_eat}
\end{figure}

\begin{figure}[h]
  \centering
  \includegraphics[width=\linewidth]{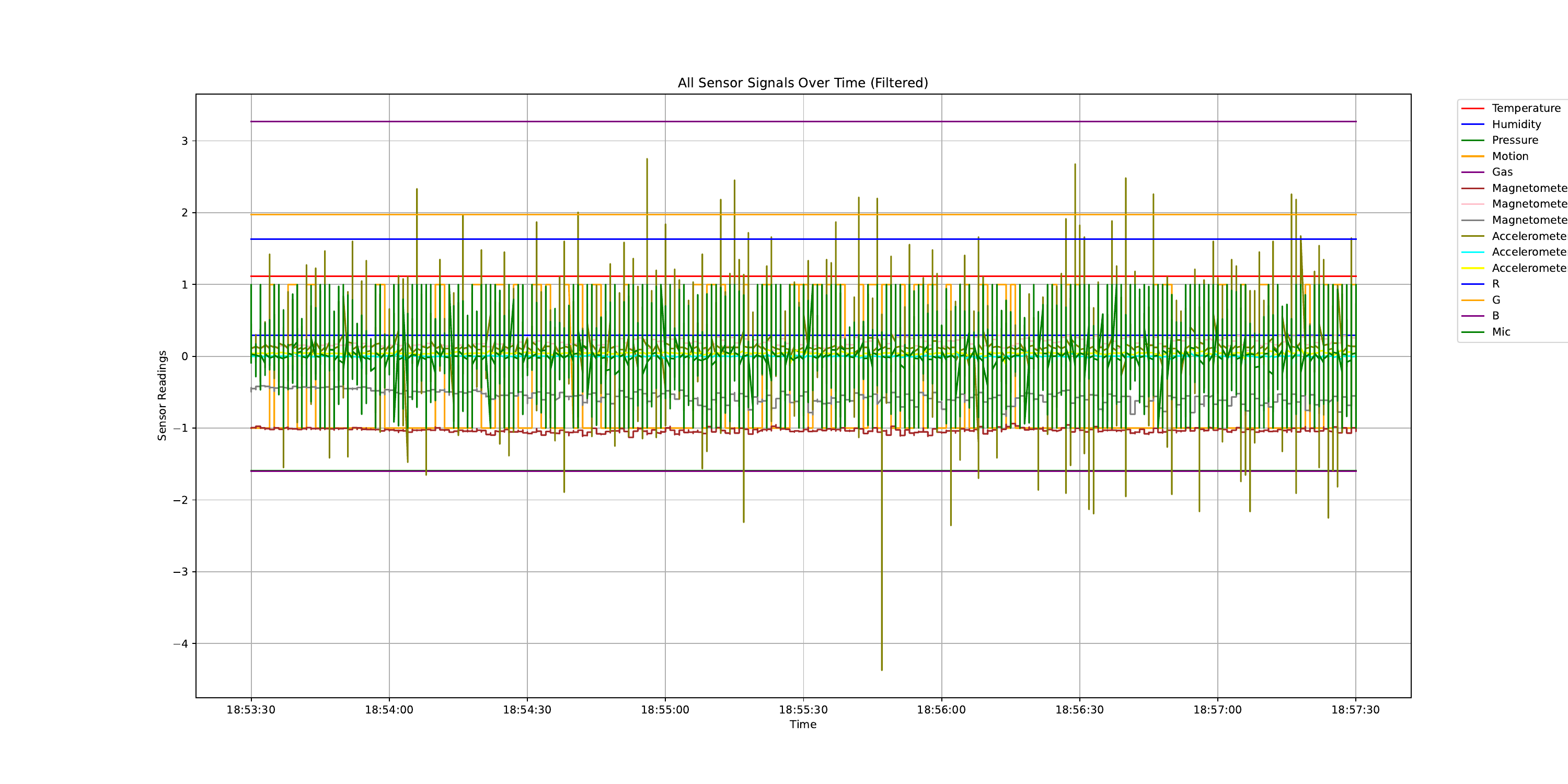}
  \caption{Data collected from chopping activities}
  \label{data_chop}
\end{figure}

The atomic level activities we collect include 20 activities: eat, paperdis, write, chop, hand wash, pour water, clean floor, knock, run, curtain, light switch, type, door pass, wipe desk, chat, basketball, saw, shave, wash dish, and brusing teeth. The sampling frequency is 90Hz for each sensor channel. There are discussions about whether the high-fidelity audio sensor channel~\cite{laput2017synthetic} is the main channel for detecting patterns. Audio data can also be intrusive~\cite{bandodkar2014non,susnea2019unobtrusive}. To avoid privacy issues associated with the audio channel, its frequency is decreased from 16kHz to 90kHz. According to Shannon’s sampling theorem, information contained in frequencies greater than half the sampling rate cannot be recovered~\cite{niazi2017statistical,khan2016optimising,banos2014window}. Figure~\ref{data_eat} visualizes the sensor data collected during eating activities. Figure~\ref{data_chop} visualizes the data collected during chopping activities.

\subsection{Initial Results}

We developed a data encoder to process the sensor data before sending it to the LLM. Our model primarily uses a channel-wise MLP to extract features from each channel, followed by a sensor fusion module also based on MLP. Additionally, we incorporated a fast Fourier convolution module~\cite{chi2020fast} to improve the accuracy rate.

Table~\ref{tab:1} presents the initial results of F1 scores, precision, and recall for detecting various activities using the model. Most activities, such as 'eat', 'write', 'chop', 'hand wash', 'clean floor', 'knock', 'run', 'curtain', 'light switch', 'type', 'door pass', 'wipe desk', 'basketball', 'saw', 'shave', 'wash dish', and 'teeth', achieve perfect scores of 1.00 across all metrics. However, 'paperdis', 'pour water', and 'chat' show lower performance with F1 scores of 0.43, 0.48, and 0.32, respectively. These variations indicate that while the model performs exceptionally well for most activities, there is room for improvement in detecting certain activities. 

\begin{figure}[h]
  \centering
  \includegraphics[width=\linewidth]{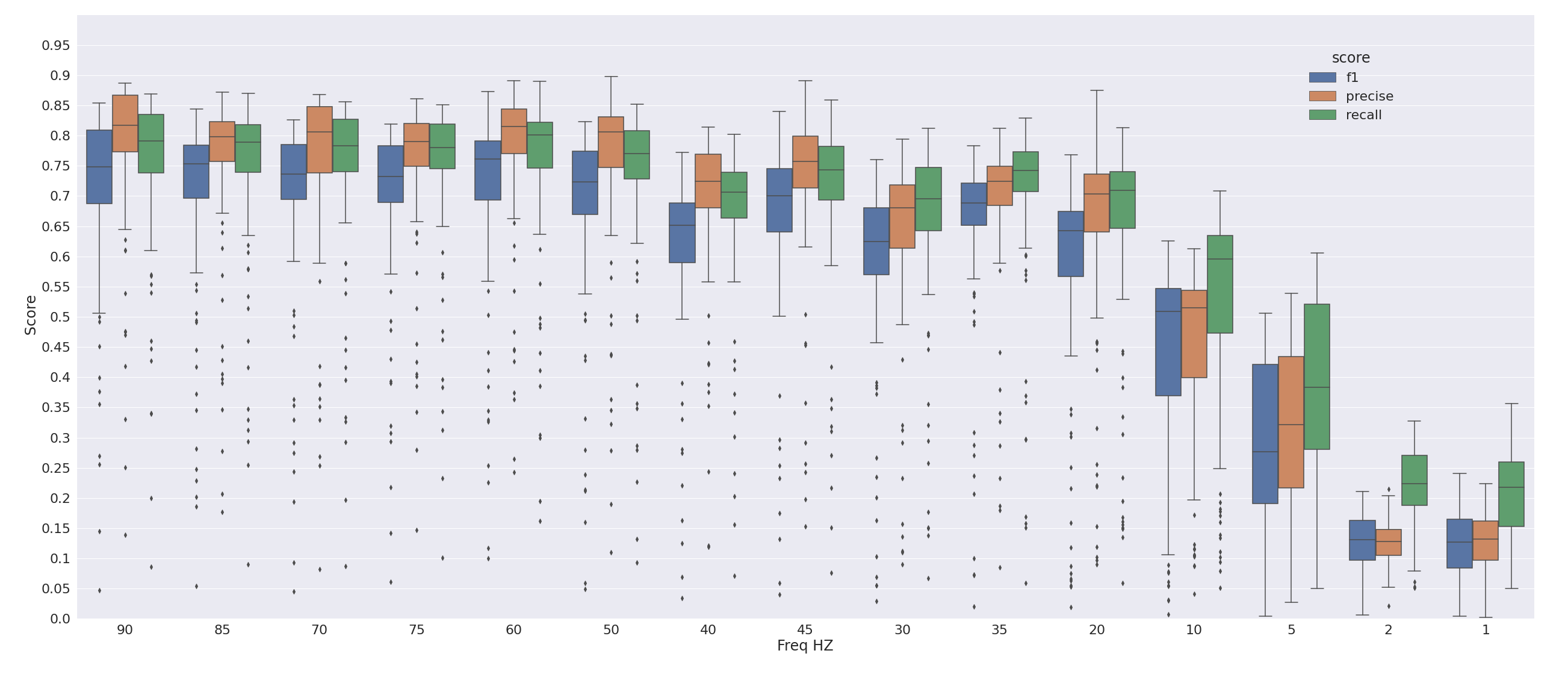}
\caption{Model robustness in detecting data at different frequencies}
  \label{lowfreq}
\end{figure}

\begin{figure}[h]
  \centering
  \includegraphics[width=\linewidth]{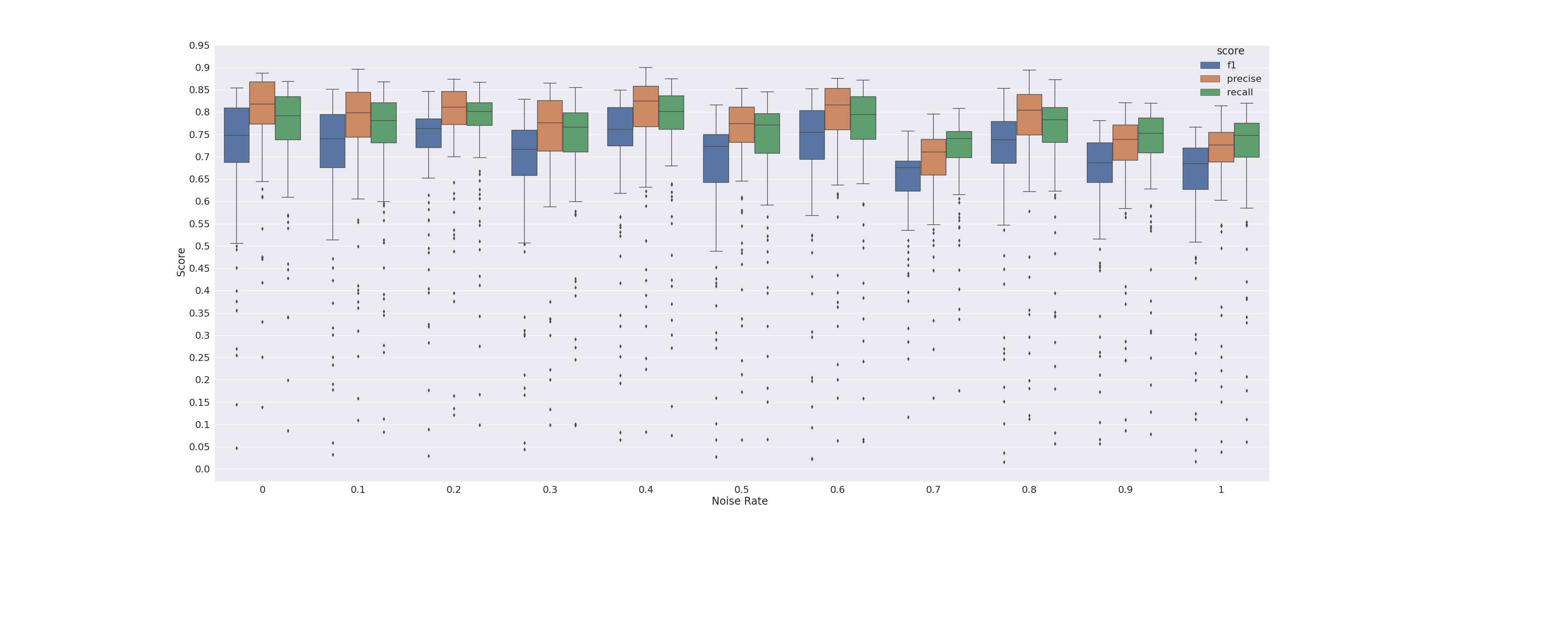}
\caption{Model robustness with varying levels of noise in the sensor data}
  \label{noiselevel}
\end{figure}

We also demonstrate the robustness of our model when the noise level~\ref{noiselevel} and frequency\ref{lowfreq} of the data vary.

\begin{table}[t]
  %\caption{Activity analysis }
  %\label{tab:3}
  \begin{tabular}{cccl}
    \toprule
    Activity Name&F1 &Precision &Recall \\
    \midrule
    
 eat &	 1.00 &	 1.00 	& 1.00 \\
 paperdis &		 0.43 &		 0.50 	&	 0.38 \\
 write 	&	 1.00 	&	 1.00 &		 1.00 \\
 chop 	 &	1.00 	&	 1.00 	&	 1.00 \\
 hand wash &		 1.00 	&	 1.00 &		 1.00 \\
 pour water &	 0.48 	&	 0.50 &		 0.47 \\
 clean floor 	&	 1.00 	&	 1.00 &		 1.00 \\
 knock 	&	 1.00 &		 1.00 &		 1.00 \\
 run &		 1.00 	&	 1.00 	&	 1.00 \\
 curtain 	&	 1.00 	&	 1.00 	&	 1.00 \\
 light Switch 	&	 1.00 &		 1.00 	&	 1.00 \\
 type &		 1.00 &		 1.00 	&	 1.00 \\
 door pass &		 1.00 	&	 1.00 	&	 1.00 \\
 wipe desk 	&	 1.00 	&	 1.00 	&	 1.00 \\
 chat 	&	 0.32 &		 0.33 &		 0.31 \\
 basketball 	&	 1.00 &		 1.00 	&	 1.00 \\
 saw 	&	 1.00 	&	 1.00 &		 1.00 \\
 shave 	&	 1.00 	&	 1.00 	 &	1.00 \\
 wash dish 	&	 1.00 	&	 1.00 	&	 1.00 \\
 teeth 	&	 1.00 &		 1.00 &		 1.00 \\

      \bottomrule
   \end{tabular}
\caption{Initial results of F1 scores for each activity}
\label{tab:1}
\end{table}

\subsection{Applying LLMs to Sequence Detection and Complex Activity Recognition}

After detecting atomic activities, we send the sequence of detected activities to the LLM. For example, to remind elderly people to take medication before eating, we send a sequence of activities such as "brushing teeth → hand wash → pour water → eat". The corrected sequence from the LLM is "brushing teeth → hand wash → take medication → pour water → eat", recognizing the complex activity as "forgetting medication". If the original order detected is "eat → basketball → brush teeth", the LLM will suggest "brush teeth → wash hands → eat → basketball" to ensure proper hygiene before eating, recognizing the complex activity as "unhygienic behavior". We use GPT-4 to implement sequence verification and reminders. For instance, if the input sequence is "door pass → using paper dispenser", the LLM suggests "door pass → turn the switch → paper dispenser" to remind the user to turn on the light, identifying the complex activity as "preventing slipping". The order rules can be configured in the prompt to meet detailed requirements. Our current results demonstrate an implementable framework that can help elderly people, especially those with memory issues, to live a high-quality life.

\section{Conclusion and Future Work}

In this work, we designed a framework leveraging IoT and LLMs to assist elderly people and enhance their quality of life. We employed a board with non-intrusive sensors to avoid discomfort for the elderly while monitoring their behavior to provide assistance. We implemented a low-frequency sensor sampling strategy, particularly for the microphone channel, using an unrecoverable sampling rate that cannot reproduce the original sound information. A smart IoT box was built to collect initial data for experimentation. Our initial results demonstrate that our model can successfully identify daily activities from the sensor data. We further utilized the GPT-4 interface to test the reasoning and complex activity detection capabilities, showcasing a promising framework that can assist elderly individuals.

However, the current smart sensor setup uses an older version of the Raspberry Pi. We plan to use a higher version of the Raspberry Pi to enable real-time inference on the edge device. Additionally, we will conduct user studies with more participants to interact with our system, helping us evaluate the design and implementation of the system.

%%
%% Print the bibliography
%%
\printbibliography

%%
%% If your work has an appendix, this is the place to put it.

\end{document}